\documentclass[12pt]{article}
\usepackage{graphicx,epsfig}
\usepackage{latexsym}
\usepackage{amsfonts,amssymb}
\usepackage{amsmath}
\usepackage[square]{natbib}
\usepackage{bm}
\def \TL{{\mathrm{L}}}
\def \TP{{\mathrm{P}}}
\def \d{{\mathrm{d}}}

\def \pd{\partial}

\def \tl#1{\overset{\kern 1pt\circ}{#1}}
\def \TL#1{\overset{\kern -3pt \circ}{#1}}
\def \TLL#1{\overset{\kern -7pt \circ}{#1}}

\def \Bsigma{\boldsymbol{\sigma}}
\def \Bbeta{\boldsymbol{\beta}}

\def \Btau{{\boldsymbol{\tau}}}

\def \Bb{{\boldsymbol{b}}}

\def \Bu{{\boldsymbol{u}}}

\def \Bx{{\boldsymbol{x}}}

\def \BR{{\boldsymbol{R}}}

\def \FF{{\cal{F}}}

\def \Bbeta{\boldsymbol{\beta}}
\def \Balpha{\boldsymbol{\alpha}}

\def \Bu{{\boldsymbol{u}}}

\def\T{{\text T}}

\def\negenspace{\kern-1.1em}

\def\negenspaceexp{\kern-0.5em}

\begin{document}

\title{{\bf
Micromechanics and dislocation theory 
in anisotropic elasticity}}
\author{
Markus Lazar~$^\text{}$\footnote{
{\it E-mail address:} lazar@fkp.tu-darmstadt.de (M.~Lazar).}
\\ \\
${}^\text{}$
        Heisenberg Research Group,\\
        Department of Physics,\\
        Darmstadt University of Technology,\\
        Hochschulstr. 6,\\
        D-64289 Darmstadt, Germany\\
}

\date{\today}
\maketitle

\begin{abstract}
In this work, dislocation master-equations valid for anisotropic materials 
are derived in terms of kernel functions 
using the framework of micromechanics. 
The second derivative of the anisotropic
Green tensor is calculated in the sense of generalized functions 
and decomposed into a sum of a $1/R^3$-term plus a Dirac $\delta$-term.
The first term is the so-called ``Barnett-term'' and 
the latter is important for the definition of the Green tensor as
fundamental solution of the Navier equation. 
In addition, all dislocation master-equations are specified for Somigliana 
dislocations with application to 3D crack modeling. 
Also the interior Eshelby tensor for a spherical inclusion in an anisotropic
material is derived as line integral over the unit circle. 
\\

\\
\end{abstract}

\section{Introduction}
The fields of defects such as dislocations, cracks, and inclusions 
play an important role in determining the physical properties of 
solids. 
Especially, dislocations play an important role 
as the elementary defect causing plasticity
and hardening in crystals \citep{Kroener58,Lardner}.
The micromechanics of isotropic inclusions and anisotropic inclusions 
based on Green tensors is an active and important research field of
the so-called eigenstrain theory~\citep{Mura,Burya,LiS}. 
Another application of the eigenstrain theory is the dislocation theory 
\citep{Mura,LiS}. 
Nowadays, the dislocation fields are important for 
computer simulations of dislocation dynamics, 
dislocation based plasticity, and dislocation based fracture mechanics. 
The classical field theory of dislocations (theory of internal stresses) 
is the theory of incompatible
anisotropic elasticity where a prescribed plastic distortion plays the role 
of the eigendistortion. 
In dislocation theory, 
the nonuniform plastic distortion gives rise to a measurable amount of incompatibility called
the dislocation density tensor \citep{Kroener58,Kroener81,Lardner}.
The prescribed plastic distortion and dislocation density tensors are the
sources for the elastic fields of dislocations. 
In the framework of field theory of anisotropic elasticity, 
the Green tensor and its derivatives play an essential role to determine the
elastic fields caused by a dislocation in an anisotropic body.

In this work, we provide anisotropic 
dislocation master-formulas based on the an\-isotropic Green tensor 
and related tensors for arbitrary dislocation
densities and plastic distortions including Volterra and Somigliana dislocations
in the framework of eigenstrain theory.
Volterra dislocations are realized as dislocations in crystals 
where the Burgers vector is fixed by the crystal lattice. 
Nowadays, Somigliana dislocations play a role as crack dislocations 
where the displacement jump (Burgers vector) is variable 
\citep{Eshelby82,Hills}. 
One aim of this work is to give a necessary set of dislocation master-equations 
for Somigliana dislocations in an anisotropic medium of infinite extent 
needed for numerical simulations of three-dimensional (3D) crack modeling.

The paper is organized into six sections. 
In Sec.~2, microelasticity and anisotropic elasticity 
with the Green tensors, their derivatives, 
related kernel functions, related tensors, the dislocation master-equations
for arbitrary prescribed plastic distortion, and dislocation density 
are introduced and derived. In Sec.~3, the dislocation master-equations are 
specified for Somigliana dislocations. In Sec.~4, the relation to 3D crack
modeling based on Somigliana dislocations is given and reviewed. In Sec.~5, 
based on the second derivative of the anisotropic Green tensor,
the Eshelby tensor for a spherical inclusion in an anisotropic material 
is given. Finally, in Sec.~6, the conclusions are given. 
Technical details are given in two Appendices.

\section{Dislocations in anisotropic elasticity }
In this Section, we derive all the necessary dislocation master-equations for arbitrary 
plastic distortion and dislocation density tensors in an anisotropic material
using the eigenstrain theory.
We use the Green tensor method since it provides a convenient framework
to calculate the fields of dislocations in anisotropic media. 
We consider linear elastic anisotropic bodies which are infinite in extent.

\subsection{Basic framework}
First of all, we review the general dislocation master-equations 
which are valid in linear anisotropic elasticity as well as in isotropic
elasticity. 
The static equilibrium condition for self-stresses reads
\begin{align}
\label{EC}
\sigma_{ij,j}=0\,,
\end{align}
where $\Bsigma$ is the symmetric stress tensor.
We use the indicial comma notation to indicate spatial differentiation: 
$\pd_j$ is indicated by the subscript notation ``$,j$''.
For the general anisotropic case, the stress tensor
is related to the elastic distortion tensor $\Bbeta$ by Hooke's law
\begin{align}
\label{HL}
\sigma_{ij}=C_{ijkl}\beta_{kl}\,,
\end{align}
where $C_{ijkl}$ is the fourth-rank tensor of elastic constants.
The tensor $C_{ijkl}$ satisfies the symmetry conditions
\begin{align}
\label{C-sym}
C_{ijkl}=C_{klij}=C_{jikl}=C_{ijlk}\,.
\end{align}
According to \citet{Kroener58}, 
the gradient of the displacement vector $\bm u$, which is the total distortion
tensor, can be decomposed into the elastic distortion tensor $\Bbeta$ and 
the plastic distortion tensor (or eigendistortion tensor) $\Bbeta^\TP$ 
\begin{align}
\label{B-deco}
u_{i,j}=\beta_{ij}+\beta_{ij}^\TP\,. 
\end{align}

The dislocation density tensor $\Balpha$ is defined in terms of the 
plastic distortion tensor~\citep{Kroener58,Kroener81}
\begin{align}
\label{A-P}
\alpha_{ij}=-\epsilon_{jkl}\beta^{\TP}_{il,k}\,
\end{align}
or in terms of the elastic distortion tensor
\begin{align}
\label{A-el}
\alpha_{ij}=\epsilon_{jkl}\beta_{il,k}\,,
\end{align}
where $\epsilon_{jkl}$ is the Levi-Civita tensor or permutation tensor. 
The dislocation density tensor satisfies the dislocation Bianchi identity
\begin{align}
\alpha_{ij,j}=0\,,
\end{align}
which has the meaning that dislocations cannot end inside the body.

\subsection{General dislocation master-equations}

If we substitute Eqs.~(\ref{HL}) and (\ref{B-deco}) into
Eq.~(\ref{EC}), we obtain an inhomogeneous Navier equation for 
the displacement vector
\begin{align}
\label{u-Navier}
C_{ijkl} u_{k,lj}=C_{ijkl} \beta_{kl,j}^\TP\,,
\end{align}
where the inhomogeneous part (or source part) 
is given by the plastic distortion tensor $\Bbeta^\TP$.
The Green tensor of the Navier equation 
(elliptic partial differential equation of second order) 
is defined by
\begin{align}
\label{GT}
C_{ijkl}G_{km,lj}(\Bx-\Bx')+\delta_{im}\delta(\Bx-\Bx')=0\,
\end{align}
and is the fundamental solution of the Navier operator 
$C_{ijkl}\pd_j\pd_l$.
$\delta_{im}$ is the Kronecker delta tensor and 
$\delta(\Bx-\Bx')$ is the three-dimensional Dirac delta function
which is zero  everywhere, except at point $\Bx=\Bx'$.
The Green tensor $G_{ij}(\Bx-\Bx')$
satisfies the symmetry relations~\citep{Bacon}
\begin{align}
\label{GT-sym}
G_{ij}(\Bx-\Bx')= G_{ij}(\Bx'-\Bx)= G_{ji}(\Bx-\Bx')\,
\end{align}
and 
\begin{align}
\label{GT-sym1}
&G_{ij,k}(\Bx-\Bx')= -G_{ij,k'}(\Bx-\Bx')= -G_{ij,k}(\Bx'-\Bx)\\
\label{GT-sym2}
&G_{ij,kl}(\Bx-\Bx')= G_{ij,k'l'}(\Bx-\Bx')= -G_{ij,kl'}(\Bx-\Bx')=
G_{ij,kl}(\Bx'-\Bx)\,,
\end{align}
where the unprimed subscripts denote partial differentiation with respect 
to $\Bx$ and primed subscripts denote differentiation with respect to $\Bx'$.
Using the Green tensor, the solution of Eq.~(\ref{u-Navier}) is the 
convolution of the Green tensor  with the plastic distortion tensor $\Bbeta^\TP$
\begin{align}
\label{u-sol}
u_i=-C_{jkln} G_{ij,k}*\beta_{ln}^\TP\,,
\end{align}
where $*$ denotes the spatial convolution.
Eq.~(\ref{u-sol}) is the generalized Volterra formula for an arbitrary
plastic distortion. 

The gradient of Eq.~(\ref{u-sol}) gives
\begin{align}
\label{grad-u}
u_{i,m}&=-C_{jkln} G_{ij,km}*\beta_{ln}^\TP\,.
\end{align}
The elastic distortion is obtained from Eq.~(\ref{B-deco}) as
\begin{align}
\label{B-1}
\beta_{im}=-C_{jkln} G_{ij,km}*\beta_{ln}^\TP-\beta_{im}^\TP\,.
\end{align}
Defining the kernel $\bm R$~\citep{Simmons}
\begin{align}
\label{R}
R_{imln}(\Bx-\Bx')&:=-C_{jkln} G_{ij,km'}(\Bx-\Bx')+\delta_{il}\delta_{mn}\delta(\Bx-\Bx')
\nonumber\\
&\ =C_{jkln} G_{ij,km}(\Bx-\Bx')+\delta_{il}\delta_{mn}\delta(\Bx-\Bx')\,,
\end{align}
Eq.~(\ref{B-1}) reduces to
\begin{align}
\label{B-2}
\beta_{im}=-R_{imln}*\beta_{ln}^\TP\,.
\end{align}
Using Eq.~(\ref{HL}), the stress tensor reads
\begin{align}
\label{stress-1}
\sigma_{pq}=-C_{pqim}\big(C_{jkln} G_{ij,km}*\beta_{ln}^\TP+\beta_{im}^\TP\big)\,,
\end{align}
which can be simply written
\begin{align}
\label{stress-2}
\sigma_{pq}=-S_{pqln}*\beta_{ln}^\TP\,,
\end{align}
where the kernel $\bm S$ is defined by~(see also \citet{Simmons,Kunin})
\begin{align}
\label{S}
S_{pqln}(\Bx-\Bx')&:=C_{pqim}R_{imln}(\Bx-\Bx')\nonumber\\
&\ =C_{pqim}\big(C_{jkln} G_{ij,km}(\Bx-\Bx')+\delta_{il}\delta_{mn}\delta(\Bx-\Bx')\big)\,.
\end{align}
$S_{pqln}(\Bx-\Bx')$ is symmetric in the indices $pq$ and $ln$ and possesses 
the additional symmetry properties
\begin{align}
\label{S-sym}
S_{pqln}(\Bx-\Bx')= S_{pqln}(\Bx'-\Bx)= S_{lnpq}(\Bx-\Bx')
\end{align}
following from the symmetries (\ref{C-sym}) and (\ref{GT-sym}).
Eqs.~(\ref{B-1}) and (\ref{stress-1}) give the 
elastic distortion and the stress, respectively,   
for a prescribed plastic distortion (or eigendistortion).  
Note that in elasticity, the kernels $\bm R$ and $\bm S$ possess 
$1/R^3$- and $\delta$-singularities.
Formulas like~(\ref{B-1}) and (\ref{stress-1})
are also known in the Eshelby-Mura eigenstrain theory and micromechanics
(see, e.g., \citet{Mura,LiS}).
In micromechanics, the second derivative of the Green tensor
is more often used than the Green tensor and the first derivative of the 
Green tensor.

On the other hand, Eq.~(\ref{grad-u}) can be rewritten as
\begin{align}
\label{grad-u2}
u_{i,m}&=-C_{jkln} G_{ij,k}*\beta_{ln,m}^\TP\nonumber\\
&=-C_{jkln} G_{ij,k}*\big(\beta_{lm,n}^\TP+\epsilon_{nmr}\alpha_{lr}\big)
\nonumber\\
&=-C_{jkln} G_{ij,kn}*\beta_{lm}^\TP
-\epsilon_{nmr}C_{jkln} G_{ij,k}*\alpha_{lr}
\nonumber\\
&=\beta_{im}^\TP
+\epsilon_{mnr}C_{jkln} G_{ij,k}*\alpha_{lr}\,,
\end{align}
which gives now for the elastic distortion tensor the so-called
Mura-Willis formula for an arbitrary dislocation density
\begin{align}
\label{Mura1}
\beta_{im}=\epsilon_{mnr}C_{jkln} G_{ij,k}*\alpha_{lr}\,.
\end{align}
Defining the kernel $\bm D$~\citep{Simmons}
\begin{align}
\label{D}
D_{imlr}(\Bx-\Bx'):=\epsilon_{mnr}C_{jkln} G_{ij,k}(\Bx-\Bx')
\end{align}
which obeys the symmetry properties
\begin{align}
\label{D-sym}
D_{imlr}(\Bx-\Bx')= -D_{imlr}(\Bx'-\Bx)= -D_{irlm}(\Bx-\Bx')\,,
\end{align}
Eq.~(\ref{Mura1}) simply reads
\begin{align}
\label{Mura2}
\beta_{im}=D_{imlr}*\alpha_{lr}\,. 
\end{align}
Substituting Eq.~(\ref{Mura1}) into Eq.~(\ref{HL}), 
the stress tensor reduces to 
\begin{align}
\label{PKs1}
\sigma_{pq}=C_{pqim}\epsilon_{mnr}C_{jkln} G_{ij,k}*\alpha_{lr}
\end{align}
which is the generalized Peach-Koehler stress formula for an 
arbitrary dislocation density. 
Eq.~(\ref{PKs1}) can be simply written as follows
\begin{align}
\label{PKs2}
\sigma_{pq}=M_{pqlr}*\alpha_{lr}
\end{align}
if we introduce the kernel $\bm M$ 
\begin{align}
\label{M}
M_{pqlr}(\Bx-\Bx')&:=C_{pqim}D_{imlr}(\Bx-\Bx') \nonumber\\
&\ =C_{pqim}\epsilon_{mnr}C_{jkln} G_{ij,k}(\Bx-\Bx')
\end{align}
which possesses the symmetry properties
\begin{align}
\label{M-sym}
M_{pqlr}(\Bx-\Bx')= -M_{pqlr}(\Bx'-\Bx)= M_{qplr}(\Bx-\Bx')\,.
\end{align}
Note that the curl of 
the kernels $\bm D$ and $\bm M$ leads to the kernels 
$\bm R$ and $\bm S$, respectively
\begin{align}
\label{D-R}
&\epsilon_{qrp}D_{imlr,p}=R_{imlq}\\
\label{M-S}
&\epsilon_{qrp}M_{imlr,p}=S_{imlq}\,.
\end{align}
Thus, Eqs.~(\ref{Mura2}) and (\ref{PKs2}) give the 
elastic distortion and the stress, respectively,   
for a prescribed dislocation density.  
Note that in elasticity, the kernels $\bm D$ and $\bm M$ possess 
a $1/R^2$-singularity.
Thus, Eqs.~(\ref{Mura2}) and (\ref{PKs2}) are less singular than
Eqs.~(\ref{B-1}) and (\ref{stress-1}).
In dislocation theory, the first derivative of the Green tensor
is of primary importance.

From Eq.~(\ref{B-deco}), we can derive the following Poisson equation for the
displacement vector
\begin{align}
\label{u-P}
\Delta u_i=\beta_{ij,j}+\beta_{ij,j}^\TP\,,
\end{align}
where $\Delta$ is the Laplace operator.
Using the three-dimensional Green function of the Laplace operator~\citep{Wl}
\begin{align}
\label{G}
G^\Delta=-\frac{1}{4\pi R}\,,\qquad \Delta G^\Delta=\delta(\Bx-\Bx')\,,
\end{align}
where $R=|\Bx-\Bx'|$ is the distance between field and source points, 
the solution of Eq.~(\ref{u-P}) can be written as convolution 
\begin{align}
u_i=\big(\beta_{ij,j}+\beta_{ij,j}^\TP\big)*G^\Delta
\end{align}
and finally
\begin{align}
\label{u0-1}
u_i=G^\Delta_{,j}*\beta^\TP_{ij}+\beta_{ij,j}*G^\Delta\,.
\end{align}
Substituting the Mura-Willis equation~(\ref{Mura1}) into Eq.~(\ref{u0-1}),
we obtain 
\begin{align}
u_i&=G^\Delta_{,j}*\beta^\TP_{ij}+
\epsilon_{mnr}C_{jkln} G_{ij,km}*\alpha_{lr}*G^\Delta\nonumber\\
&=G^\Delta_{,j}*\beta^\TP_{ij}+
\epsilon_{mnr}C_{jkln} G_{ij,km}*G^\Delta*\alpha_{lr}\,.
\end{align}
The double convolution can be reduced to a single one by defining the 
$\bm F$-tensor introduced by~\citet{Kirchner1} (see also~\citet{LK13})
\begin{align}
\label{F}
F_{mkij}:=-G_{ij,km}*G^\Delta\,.
\end{align}
Thus, the $\bm F$-tensor is the convolution of the second derivative of the 
Green tensor with the Green function of the Laplace operator.
It satisfies a tensorial Poisson equation
\begin{align}
\label{F-pde}
\Delta F_{mkij}=-G_{ij,km}\,.
\end{align}
Moreover, the following relationships hold
\begin{align}
\label{F-rel1}
F_{mkij,m}+G_{ij,k}=0\,,\qquad
F_{mkij,k}+G_{ij,m}=0\,,
\end{align}
\begin{align}
\label{F-rel2}
F_{mkij,mk}+\Delta G_{ij}=0\,,
\end{align}
and 
\begin{align}
\label{F-tr}
\delta_{mk}F_{mkij}=-G_{ij}\,.
\end{align}
$F_{mkij}$ obeys the symmetry properties
\begin{align}
\label{F-sym}
F_{mkij}=F_{kmij}=F_{mkji}\,,
\end{align}
and 
\begin{align}
\label{F-sym2}
F_{mkij}(\Bx-\Bx')=F_{mkij}(\Bx'-\Bx)\,.
\end{align}
Combining 
Eqs.~(\ref{GT}) and (\ref{F-pde}), we obtain
\begin{align}
\label{F-rel3}
\Delta C_{iklm}F_{mkij}=-C_{iklm}G_{ij,km}=\delta_{jl}\delta(\bm r-\bm r')
\end{align}
and from (\ref{F}) with (\ref{GT}), we get
\begin{align}
\label{F-rel4}
C_{iklm}F_{mkij}=-C_{iklm}G_{ij,km}*G^\Delta=\delta_{jl}G^\Delta\,.
\end{align}
Using the $\bm F$-tensor, 
the displacement is given by the generalized Burgers equation
\begin{align}
\label{u-Burgers}
u_i&=G^\Delta_{,j}*\beta^\TP_{ij}
-\epsilon_{mnr} C_{jkln} F_{mkij}*\alpha_{lr}\,.
\end{align}
Eq.~(\ref{u-Burgers}) gives a decomposition of the displacement vector 
into a purely geometric part convoluted with a prescribed plastic distortion
and a part depending on the elastic constants convoluted with the
dislocation density tensor. 

All the given equations are valid for anisotropic elasticity as well as
isotropic elasticity. 
Only the corresponding Green tensor and $\bm F$-tensor have to be
substituted. 
They are given for any plastic distortion and dislocation density,
that means they are valid for Volterra dislocations as well as 
for Somigliana dislocations (see Sec.~\ref{S-disloc}).

\subsection{Green tensor, $\bm F$-tensor, their derivatives and corresponding 
kernels for arbitrary anisotropic elasticity}

\begin{figure}[t]
\centering
\includegraphics[width=0.35\textwidth]{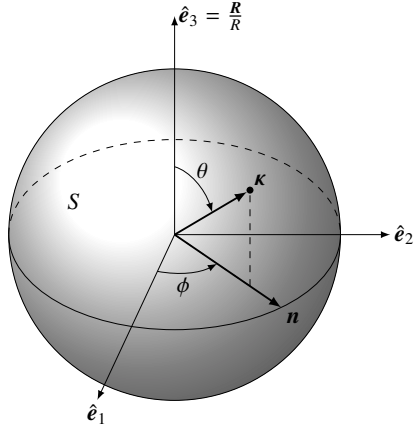}
\caption{The unit sphere in Fourier space. 
The unit vector $\bm \kappa(\theta,\phi)$ is defined by the azimuth 
angle $\phi$, and the  zenith angle $\theta$  
and its projection $\bm n=\bm \kappa(\pi/2,\phi)$
is measured from the axis $\hat{\bm e}_3=\bm R/R$.}
\label{kSpace}
\end{figure}

For anisotropic elasticity, the Green tensor reads 
\citep{LR,Synge,Barnett}
\begin{align}
\label{GT-0}
G_{ij}(\bm R)
&=\frac{1}{8\pi^2 R}\, \int_0^{2\pi} 
(n C n)_{ij}^{-1}\, \text{d} \phi\,,
\end{align}
where $\bm n=\bm \kappa(\pi/2,\phi)$ and $\bm R=\Bx-\Bx'$. 
The second rank symmetric tensor, 
sometimes called Christoffel stiffness tensor, 
is defined as 
\begin{align}
\label{nCn}
(nCn)_{ij}=n_kC_{ikjl} n_l\,
\end{align}
and the inverse tensor $(nCn)^{-1}_{ij}$ of $(nCn)_{ij}$
is defined by the property 
\begin{align}
\label{nCn-1}
(nC n)^{-1}_{ip}\,(n C n)_{pj}=
(nC n)_{ip}\,(n C n)_{pj}^{-1}=\delta_{ij}\,.
\end{align}
Here $\bm \kappa=\bm k/k$ with $k=|\bm k|$ is a unit vector 
in the Fourier space with $\bm \kappa=\bm \kappa(\theta,\phi)$. 
Eq.~(\ref{GT-0}) is the famous Lifshitz-Rosenzweig-Synge-Barnett 
representation of the Green tensor for arbitrary anisotropy
which is a line   
integral along the unit circle in the plane orthogonal to $\bm R$ 
(see Fig.~\ref{kSpace}).
Thus, $\bm n$ is perpendicular to $\bm R$, namely $\bm n\cdot\bm R=0$. 

The first derivative of the Green tensor~(\ref{GT-0}) is given by 
(see Appendix~\ref{AppendixB} and \citet{Barnett,Bacon})
\begin{align}
\label{GT-grad-1}
G_{ij,k}(\BR)=
-\frac{1}{8\pi^2 R^2}\, \int_0^{2\pi}\Big(\tau_k (n C n)_{ij}^{-1}
-n_k F_{ij}
\Big)\, \d \phi\, 
\end{align}
with the unit vector
\begin{align}
\label{tau}
\Btau=\frac{\BR}{R}
\end{align}
and
\begin{align}
\label{F-ij}
F_{ij}=(n C n)_{ip}^{-1}\big[(n C \tau)_{pq}+(\tau C n)_{pq}\big](n C
n)_{qj}^{-1}\,.
\end{align}
The $\bm F$-tensor reads~\citep{Kirchner1,LK13}
\begin{align}
\label{F-0}
F_{mkij}(\bm R)
&=-\frac{1}{8\pi^2 R}\, \int_0^{2\pi} n_m n_k\,(n C n)_{ij}^{-1}\,  \text{d} \phi\, ,
\end{align}
which is also a
line integral along the unit circle in the plane orthogonal to $\bm R$.
Using Eqs.~(\ref{F-0}) and (\ref{nCn-1}), it can be easily seen that the relation~(\ref{F-rel4})
is fulfilled, and using  Eqs.~~(\ref{F-0}), (\ref{nCn-1}) and (\ref{G}),  
it can be easily seen that the relation~(\ref{F-rel3}) is satisfied.

An important similarity is that both $G_{ij}(\bm R)$ and $F_{mkij}(\bm R)$ 
have a $1/R$-singularity and their Fourier transforms $\hat G_{ij}(\bm k)$ 
and $\hat F_{mkij}(\bm k)$ are proportional to $k^{-2}$. 
On the other hand, the second derivative of the Green tensor,
$G_{ij,km}(\bm R)=-\Delta F_{mkij}(\bm R)$, is a generalized 
function (distribution), whose Fourier transform, 
$k_k k_m \hat G_{ij}(\bm k)=-k^2 \hat F_{mkij}(\bm k)$, 
is a homogeneous function of zeroth degree. 
From the theory of generalized functions~\citep{GS}, it follows that the inverse 
Fourier transform, 
$G_{ij,km}(\bm R)=-\Delta F_{mkij}(\bm R)$, of the generalized homogeneous 
function of zeroth degree, 
$k^2 \hat F_{mkij}(\bm k)=-k_k k_m \hat G_{ij}(\bm k)$,
can be decomposed into a regular distribution and a singular distribution, 
which is proportional to $\delta(\bm R)$ (see also~\citet{Kunin}). 
Thus, the second derivative (second gradient) of the Green tensor can be decomposed into two terms, namely (see also~\citet{Kunin,Kroener90})
\begin{align}
\label{gradgradG-deco}
G_{ij,km}=-\delta(\bm R) E_{ijkm}+\frac{1}{R^3}\, H_{ijkm}\,.
\end{align}
The first term in Eq.~(\ref{gradgradG-deco}) is due to the derivative in the
sense of generalized functions,
and the second term is due to the formal (or ordinary) derivative of $G_{ij}$. 

Using the $\bm F$-kernel~(\ref{F-0}) and Eqs.~(\ref{F-pde}) and (\ref{G}), 
we have 
\begin{align}
\label{GT-grad-2-rel}
G_{ij,km}(\bm R)&=-\Delta F_{mkij}(\bm R)\nonumber\\
&=-\frac{1}{8\pi^2 }\, 
\Big(4\pi\delta(\bm R) +2\frac{\tau_l}{R^2}\pd_l -\frac{1}{R}\, \Delta \Big)
\int_0^{2\pi} n_m n_k\,(n C n)_{ij}^{-1}\,  \text{d} \phi\, .
\end{align}
After a straightforward calculation (see Appendix~\ref{AppendixB}), 
we obtain for the second derivative of the 
Green tensor
the following decomposition into a Dirac $\delta(\bm R)$-term and a $1/R^3$-term
\begin{align}
\label{GT-grad-2}
G_{ij,km}(\bm R)
&=-\frac{1}{2\pi}\, \delta(\bm R) \int_0^{2\pi} 
n_m n_k\,(n C n)_{ij}^{-1}\,  \text{d} \phi\nonumber\\
&\quad +\frac{1}{8\pi^2 R^3} \int_0^{2\pi} \Big(
2\tau_m \tau_k\,(n C n)_{ij}^{-1}
-2 (n_m \tau_k+n_k\tau_m) F_{ij}
+n_m n_k A_{ij}\Big)\, \text{d} \phi\,,
\end{align}
where
\begin{align}
\label{A-ij}
A_{ij}&=F_{ip}\big[(n C \tau)_{pq}+(\tau C n)_{pq}\big](n C n)_{qj}^{-1}
+(n C n)_{ip}^{-1}\big[(n C \tau)_{pq}+(\tau C n)_{pq}\big]F_{qj}
\nonumber\\
&\quad 
-2 (n C n)_{ip}^{-1}(\tau C \tau)_{pq}(n Cn)_{qj}^{-1}\,.
\end{align}
The $1/R^3$-term in Eq.~(\ref{GT-grad-2}) is the expression given 
by~\citet{Barnett}.
By construction, we obtained in Eq.~(\ref{GT-grad-2}) a decomposition 
into a singular $\delta(\bm R)$-term
and a ``regular distribution'' term proportional to $1/R^3$.
The latter is the ``Barnett-term''.  
Thus, we have in the decomposition~(\ref{gradgradG-deco}) 
of Eq.~(\ref{GT-grad-2}) the two tensors
\begin{align}
\label{E-aniso}
E_{ijkm}=\frac{1}{2\pi}\,\int_0^{2\pi} 
n_m n_k\,(n C n)_{ij}^{-1}\,  \text{d} \phi
\end{align}
and 
\begin{align}
\label{H-aniso}
H_{ijkm}=\frac{1}{8\pi^2} \int_0^{2\pi} \Big(
2\tau_m \tau_k\,(n C n)_{ij}^{-1}
-2 (n_m \tau_k+n_k\tau_m) F_{ij}
+n_m n_k A_{ij}\Big)\, \text{d} \phi\,.
\end{align}
Here both tensors~(\ref{E-aniso}) and (\ref{H-aniso}) are 
line integrals along the unit circle in the plane orthogonal to $\bm R$.
The line integral representation of the tensor $E_{ijkm}$ is a straightforward 
consequence of the Lifshitz-Rosenzweig-Synge-Barnett representation for the
anisotropic Green tensor and of the line integral representation for the tensor $F_{mkij}$. 
The tensor~(\ref{E-aniso}) 
corresponds to the $A$-term with an integration over the unit circle given 
by~\citet{Kunin}.
\footnote{Note that \citet{Kneer} (see also \citet{Kroener86}) 
gave a more complicated 
expression as integral over the unit sphere for the tensor $E_{ijkm}$, namely
\begin{align*}
E_{ijkm}=\frac{1}{4\pi}\,\int_0^{\pi} \text{d} \theta \sin\theta \int_0^{2\pi} 
\kappa_m \kappa_k\,(\kappa C \kappa)_{ij}^{-1}\,  \text{d} \phi\,,
\end{align*}
since \citet{Kneer} has not used the 
Lifshitz-Rosenzweig-Synge-Barnett representation for the anisotropic Green tensor.} The expressions~(\ref{F-ij}) and (\ref{A-ij}) are identical to the ones 
originally given by~\citet{Barnett}.
It can be concluded that the $\bm F$-tensor is useful for 
the calculation of the second derivative of the Green tensor 
of anisotropic elasticity by means of Eq.~(\ref{GT-grad-2-rel}). 
Although mentioned by~\citet{Kroener90} that a 
decomposition~(\ref{gradgradG-deco}) is valid in anisotropic elasticity 
nowhere the explicit form of the decomposition in anisotropic elasticity 
with the tensors~(\ref{E-aniso}) and (\ref{H-aniso}) 
has been reported before in the literature.

Let us now prove that Eq.~(\ref{GT-grad-2}) satisfies the Navier equation~(\ref{GT}) for the definition of the Green tensor
\begin{align}
\label{gradgradG-deco-2}
C_{pkim}G_{ij,km}(\bm R)=-\delta(\bm R)\, C_{pkim} E_{ijkm}+\frac{1}{R^3}\, 
C_{pkim} H_{ijkm}=-\delta(\bm R) \delta_{pj}\,.
\end{align}
The proof is as follows.
Using the decomposition~(\ref{gradgradG-deco-2}) and 
the tensors~(\ref{E-aniso}) and (\ref{H-aniso}), it follows that
\begin{align}
\label{pr1}
C_{pkim} E_{ijkm}= \frac{1}{2\pi}\,\int_0^{2\pi} 
(nC n)_{pi}\,(n C n)_{ij}^{-1}\,  \text{d} \phi
=\delta_{pj}\,,
\end{align}
due to Eq.~(\ref{nCn-1}).
On the other hand, it yields
\begin{align}
\label{pr2}
C_{pkim} H_{ijkm}&= \frac{1}{8\pi^2} \int_0^{2\pi} \Big(
2(\tau C \tau)_{pi}\,(n C n)_{ij}^{-1}
-2 \big((nC \tau)_{pi}+(\tau C n)_{pi}\big) F_{ij}
+(n C n)_{pi} A_{ij}\Big)\, \text{d} \phi\nonumber\\
&=0\,.
\end{align}
The tensor $E_{ijkm}$ guarantees that the 
second derivatives of $G_{ij}$ satisfy the definition of 
the Green tensor~(\ref{GT}). 
Thus, Eq.~(\ref{GT-grad-2}) is the correct expression for 
the second derivative of the Green tensor 
in the sense of generalized functions
and Eq.~(\ref{GT}) is proved for the expression~(\ref{GT-grad-2}).

On the other hand, this means that the expression for the second derivative 
of the Green tensor given by~\citet{Barnett} (e.g. \citet{Bacon,Mura,Teodosiu}) 
fulfills only a homogeneous Navier equation
\begin{align}
\label{gradgradG-deco-3}
C_{pkim}G_{ij,km}=0\,
\end{align}
and it does not satisfy Eq.~(\ref{GT}).

Quite analogously, in the sense of generalized functions 
the kernels $\bm R$ and $\bm S$ can be decomposed into
a Dirac $\delta$-part and a $1/R^3$-part.
Substituting Eq.~(\ref{GT-grad-2}) into Eq.~(\ref{R}), we obtain 
the kernel $\bm R$ for an arbitrary anisotropic medium
\begin{align}
\label{R-2}
R_{imln}(\Bx-\Bx')&=
\frac{1}{8\pi^2 R^3} \int_0^{2\pi} 
C_{jkln}
\Big(2\tau_m \tau_k\,(n C n)_{ij}^{-1}
-2 (n_m \tau_k+n_k\tau_m) F_{ij}
+n_m n_k A_{ij}\Big)\, \text{d} \phi\nonumber\\
&\quad
+\delta(\Bx-\Bx') 
\bigg(\delta_{il}\delta_{mn}
-\frac{1}{2\pi}
\int_0^{2\pi} 
C_{jkln}
n_m n_k\,(n C n)_{ij}^{-1}\,  \text{d} \phi\bigg)\,,
\end{align}
which possesses a $1/R^3$-singularity (regular distribution part) and a Dirac $\delta$-singularity (singular distribution part).
Substituting Eq.~(\ref{GT-grad-2}) into Eq.~(\ref{S}), we obtain 
the kernel $\bm S$ for an arbitrary anisotropic medium
\begin{align}
\label{S-2}
S_{pqln}(\Bx-\Bx')&=
\frac{1}{8\pi^2 R^3} \int_0^{2\pi} 
C_{pqim}C_{jkln}
\Big(2\tau_m \tau_k\,(n C n)_{ij}^{-1}
-2 (n_m \tau_k+n_k\tau_m) F_{ij}
+n_m n_k A_{ij}\Big)\, \text{d} \phi\nonumber\\
&\quad
+\delta(\Bx-\Bx') 
\bigg(C_{pqln}
-\frac{1}{2\pi}
\int_0^{2\pi} 
C_{pqim} C_{jkln}
n_m n_k\,(n C n)_{ij}^{-1}\,  \text{d} \phi\bigg)\,,
\end{align}
which possesses a $1/R^3$-singularity (regular distribution part) and a Dirac $\delta$-singularity (singular distribution part).
Substituting Eq.~(\ref{GT-grad-1}) into Eq.~(\ref{D}), we obtain 
the kernel $\bm D$ for an arbitrary anisotropic medium
\begin{align}
\label{D-2}
D_{imlr}(\Bx-\Bx')=
-\frac{1}{8\pi^2 R^2}\, \int_0^{2\pi}
\epsilon_{mnr}C_{jkln} 
\Big(\tau_k (n C n)_{ij}^{-1}
-n_k F_{ij}
\Big)\, \d \phi\, ,
\end{align}
which possesses a $1/R^2$-singularity.
Substituting Eq.~(\ref{GT-grad-1}) into Eq.~(\ref{M}), we obtain 
the kernel $\bm M$ for an arbitrary anisotropic medium
\begin{align}
\label{M-2}
M_{pqlr}(\Bx-\Bx')=
-\frac{1}{8\pi^2 R^2}\, \int_0^{2\pi}
C_{pqim}\epsilon_{mnr}C_{jkln} 
\Big(\tau_k (n C n)_{ij}^{-1}
-n_k F_{ij}
\Big)\, \d \phi\, ,
\end{align}
which possesses a $1/R^2$-singularity.

\subsection{Green tensor, $\bm F$-tensor and their derivatives 
for isotropic elasticity}

For isotropic elasticity, the Green tensor is given by (see, e.g., \citet{Mura,LiS})
\begin{align}
\label{GT-iso}
G_{ij}(\bm R)=\frac{1}{16\pi\mu(1-\nu)}\,
\big[2(1-\nu)\delta_{ij}\Delta-\pd_i\pd_j\big] R\,,
\end{align}
where $\mu$ is the shear modulus and $\nu$ is the Poisson ratio.
Using Eqs.~(\ref{Rij}), (\ref{Rii}) and (\ref{tau}), Eq.~(\ref{GT-iso}) reduces to 
\begin{align}
\label{GT-iso2}
G_{ij}(\bm R)=\frac{1}{16\pi\mu(1-\nu)}\,\frac{1}{R}
\Big[(3-4\nu)\,\delta_{ij}+\tau_i \tau_j\Big] \,.
\end{align}
Using Eqs.~(\ref{Rijk}), (\ref{Riik}) and (\ref{tau}), the first gradient of the
Green tensor is given by
\begin{align}
\label{GT-grad-iso}
G_{ij,k}(\bm R)&=-
\frac{1}{16\pi\mu(1-\nu)}\,\frac{1}{R^2}
\Big[(3-4\nu) \delta_{ij}\tau_k
-\big(\delta_{jk}\tau_i+\delta_{ik}\tau_j\big)
+3\tau_i\tau_j\tau_k\Big]\,.
\end{align}

On the other hand, the $\bm F$-tensor reads~\citep{LK13}
\begin{align}
\label{F-iso}
F_{mkij}(\bm R)=-\frac{1}{12\cdot 16\pi\mu(1-\nu)}\,
\pd_m\pd_k\big[2(1-\nu)\delta_{ij}\Delta-\pd_i\pd_j\big] R^3\,.
\end{align}
Using Eqs.~(\ref{R3ijkl}), (\ref{R3iikl}) and (\ref{tau}),
the explicit form of the $\bm F$-tensor reduces to 
\begin{align}
\label{F-iso-2}
F_{mkij}(\bm R)&=-\frac{1}{8\pi\mu(1-\nu)}\,\frac{1}{R}
\Big[(1-\nu)\delta_{ij}
\big(\delta_{km}-\tau_k \tau_m \big)\nonumber\\
&\qquad
-\frac{1}{8}\Big(\delta_{ijkm}
-\big(\delta_{ij}\tau_k \tau_m+\delta_{ik}\tau_j \tau_m+\delta_{im}\tau_j \tau_k
+\delta_{jk}\tau_i \tau_m+\delta_{jm} \tau_i \tau_k+\delta_{km}\tau_i\tau_j\big)
\nonumber\\&\qquad\qquad\quad
+3 \tau_i\tau_j \tau_k\tau_m\Big)\Big]
\,. 
\end{align} 
Using Eqs.~(\ref{F-pde}) and (\ref{F-iso}),
we obtain for the second gradient of the Green tensor  
\begin{align}
\label{GT-grad2-iso1}
G_{ij,km}(\bm R)=-\Delta F_{mkij}(\bm R)
=\frac{1}{16\pi\mu(1-\nu)}\,
\pd_m\pd_k\big[2(1-\nu)\delta_{ij}\Delta-\pd_i\pd_j\big] R\,
\end{align}
since $\Delta R^3=12 R$. 
By the help of Eqs.~(\ref{Rijkl}), (\ref{Riikl}) and (\ref{tau}), 
we obtain the explicit form of the second gradient of the Green tensor,
namely the following decomposition into a $\delta(\bm R)$-term and a $1/R^3$-term
\begin{align}
\label{GT-grad2-iso2}
G_{ij,km}(\bm R)&=-\frac{1}{3\mu}\, \delta(\bm R)
\Big[\delta_{ij}\delta_{km}-\frac{1}{10(1-\nu)}\,\delta_{ijkm}\Big]\nonumber\\
&\quad
+\frac{1}{16\pi\mu(1-\nu)}\,\frac{1}{R^3}
\Big[4(1-\nu)\delta_{ij}\big(3\tau_k\tau_m-\delta_{km}\big)
+\delta_{ijkm}+15\tau_i\tau_j\tau_k\tau_m
\nonumber\\
&\qquad
-3\big(
\delta_{ij}\tau_k\tau_m+\delta_{ik}\tau_j\tau_m+\delta_{im}\tau_j\tau_k
+\delta_{jk}\tau_i\tau_m+\delta_{jm}\tau_i\tau_k+\delta_{km}\tau_i\tau_j
\big)
\Big]\,,
\end{align}
where 
\begin{align}
\label{delta-4}
\delta_{ijkm}=\delta_{ij}\delta_{km}+\delta_{ik}\delta_{jm}+\delta_{im}\delta_{jk}\,.
\end{align}
Using the decomposition~(\ref{gradgradG-deco}), Eq.~(\ref{GT-grad2-iso2}) 
can be decomposed into the two tensors (see also~\citet{Burya})
\begin{align}
\label{E-iso}
E_{ijkm}&=\frac{1}{3\mu}
\Big[\delta_{ij}\delta_{km}-\frac{1}{10(1-\nu)}\,\delta_{ijkm}\Big]\\
\label{H-iso}
H_{ijkm}&=\frac{1}{16\pi\mu(1-\nu)}
\Big[4(1-\nu)\delta_{ij}\big(3\tau_k\tau_m-\delta_{km}\big)
+\delta_{ijkm}+15\tau_i\tau_j\tau_k\tau_m
\nonumber\\
&\qquad\qquad
-3\big(
\delta_{ij}\tau_k\tau_m+\delta_{ik}\tau_j\tau_m+\delta_{im}\tau_j\tau_k
+\delta_{jk}\tau_i\tau_m+\delta_{jm}\tau_i\tau_k+\delta_{km}\tau_i\tau_j
\big)
\Big]\,.
\end{align}

\subsection{Peach-Koehler force}

The Peach-Koehler force is defined as (see, e.g., \citet{LK13,AL10})
\begin{align}
\label{PK}
\FF_k^{\text{PK}}
= \int_S \big[W\delta_{jk}-\sigma_{ij} \beta_{ik}\big] \d S_j
=\int_V \epsilon_{kjl} \sigma_{ij}\alpha_{il}\, \d V\,,
\end{align}
with the elastic strain energy density
\begin{align}
\label{W}
W=\frac{1}{2}\, \sigma_{ij}\beta_{ij}\,,
\end{align} 
the Eshelby stress tensor~\citep{Eshelby75} in the bracket on the left of Eq.~(\ref{PK})
\begin{align}
\label{EST}
P_{kj}=W\delta_{jk}-\sigma_{ij} \beta_{ik}
\end{align}
and $S$ is the boundary surface of the volume $V$.
The Peach-Koehler force~(\ref{PK}) is the interaction force between 
a dislocation density~$\alpha_{il}$ and a stress field $\sigma_{ij}$. 

\section{Somigliana dislocation}
\label{S-disloc}

\begin{figure}[t]\unitlength1cm
\centerline{
\begin{picture}(6,5)
\put(0.0,0.0){\epsfig{file=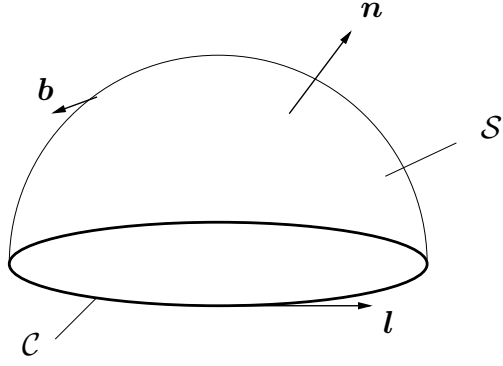,width=6cm}}
\put(5.0,0.1){$\bm l$}
\put(0.2,-0.2){$\cal C$}
\put(6.3,2.7){$\cal S$}
\put(4.7,4.3){$\bm n$}
\put(0.4,3.2){${\bm b}$}
\end{picture}
}
\caption{Somigliana dislocation.}
\label{fig:Loop}
\end{figure}

A general Somigliana dislocation is determined as a 
surface (dislocation surface) $\mathcal{S}$, on which
the displacement vector $\Bu$ has a jump
$\Bb(\Bx)$, the Burgers vector of a Somigliana dislocation, that changes arbitrary along 
$\mathcal{S}$. 
The surface $\mathcal{S}$ is the dislocation surface, which is a ``cap'' of the
dislocation line $\mathcal{C}$ (see Fig.~\ref{fig:Loop}).
Somigliana dislocations are relevant for geophysical applications~\citep{Eshelby73} and as so-called crack dislocations in
 3D dislocation based fracture mechanics~\citep{Hills,Ghoniem06}.

In the case of a Somigliana dislocation, the plastic distortion tensor is given by
\begin{align}
\label{BP-S}
\beta^{\TP}_{ij}(\Bx)=-\int_\mathcal{S} \delta(\Bx-\Bx') \, b_i(\Bx')\, \d S'_j\,.
\end{align}
Substituting Eq.~(\ref{BP-S}) into Eq.~(\ref{A-P}), we obtain for the 
dislocation density tensor of a Somigliana dislocation
\begin{align}
\label{A-S}
\alpha_{ij}(\Bx)&=\oint_\mathcal{C} \delta(\Bx-\Bx')\, b_i(\Bx')\, \d C'_j
+\int_\mathcal{S} \epsilon_{jkl}\,\delta(\Bx-\Bx') \, b_{i,k'}(\Bx')\, \d S'_l\,,
\end{align}
where $\mathcal{C}$ is the dislocation line bounding 
$\mathcal{S}$.
The dislocation density tensor of a Somigliana dislocation consists of two parts,
namely a contour integral similar to a Volterra dislocation and a surface integral 
including the gradient of the Burgers vector.

We give now the dislocation master-equations applied to Somigliana dislocations.
If we substitute the plastic distortion tensor~(\ref{BP-S}) into Eq.~(\ref{u-sol}), 
then the Volterra formula for a Somigliana dislocation reads
\begin{align}
\label{u-Vol}
u_i(\Bx)=\int_\mathcal{S} C_{jkln} G_{ij,k}(\Bx-\Bx')\, b_l(\Bx')\, \d S'_{n}\,,
\end{align}
which gives the displacement field as a surface integral.
On the other hand, if we substitute the plastic distortion tensor~(\ref{BP-S})
into Eqs.~(\ref{B-2}) and (\ref{stress-2}), the elastic distortion  tensor
\begin{align}
\label{B-Vol}
\beta_{im}(\Bx)=\int_\mathcal{S} R_{imln}(\Bx-\Bx')\, b_l(\Bx')\, \d S'_{n}\,
\end{align}
and the stress tensor of a Somigliana dislocation are obtained as surface integral
\begin{align}
\label{T-Vol}
\sigma_{pq}(\Bx)=\int_\mathcal{S} S_{pqln}(\Bx-\Bx')\, b_l(\Bx')\, \d S'_{n}\,.
\end{align}

The Mura-Willis formula of a Somigliana dislocation is obtained by 
substituting the dislocation density tensor~(\ref{A-S}) into Eq.~(\ref{Mura2})
\begin{align}
\label{Mura-S}
\beta_{im}(\Bx)=\oint_\mathcal{C} D_{imlr}(\Bx-\Bx')\, b_l(\Bx')\, \d C'_{r}
+\int_\mathcal{S} \epsilon_{rst} D_{imlr}(\Bx-\Bx')\,  b_{l,s'}(\Bx')\, \d S'_{t}
\end{align}
and the
Peach-Koehler stress formula of a Somigliana dislocation 
is obtained from Eq.~(\ref{PKs2})
\begin{align}
\label{PKs-S}
\sigma_{pq}(\Bx)=\oint_\mathcal{C} M_{pqlr}(\Bx-\Bx')\, b_l(\Bx')\, \d C'_{r}
+\int_\mathcal{S} \epsilon_{rst} M_{pqlr}(\Bx-\Bx')\,  b_{l,s'}(\Bx')\, \d S'_{t}\,.
\end{align}
Alternatively, we can derive the following generalized Burgers equation 
for a Somigliana dislocation if we substitute Eqs.~(\ref{BP-S}) and (\ref{A-S})
into Eq.~(\ref{u-Burgers})
\begin{align}
\label{u-Burgers-S}
u_i(\Bx)&=
\frac{1}{4\pi} \int_\mathcal{S}\frac{\tau_m}{R^2}\, b_i(\Bx')\, \d S'_m
-\oint_\mathcal{C}\epsilon_{mnr} C_{jkln} F_{mkij}(\Bx-\Bx')\, b_l(\Bx')\, \d C'_r
\nonumber\\
&\qquad
-\int_\mathcal{S}\epsilon_{rpq}\epsilon_{mnr} C_{jkln} F_{mkij}(\Bx-\Bx')\, b_{l,p'}(\Bx')\, \d S'_q\,.
\end{align}
An advantage of such a Burgers-like equation is the separation of a purely geometric
term depending only on the Burgers vector, 
namely the first term in Eq.~(\ref{u-Burgers-S}). 
Since the kernels $\bm R$ and $\bm S$ possess 
$1/R^3$- and Dirac $\delta$-singularities, 
the integration in Eqs.~(\ref{B-Vol}) and (\ref{T-Vol}) is hypersingular and 
not well defined.
On the other hand, the integrals in Eqs.~(\ref{u-Vol}), (\ref{Mura-S}) and
(\ref{PKs-S}) have $1/R^2$-singularities.
The integrals in 
Eq.~(\ref{u-Burgers-S}) possess $1/R^2$- and $1/R$-singularities.

If we substitute the dislocation density tensor of a Somigliana 
dislocation~(\ref{A-S}) into Eq.~(\ref{PK}), then the Peach-Koehler force 
of a Somigliana dislocation in a stress field $\bm\sigma$ reads
\begin{align}
\label{PK-S}
\FF_k^{\text{PK}}
=\oint_\mathcal{C} \epsilon_{kjl} \sigma_{ij} b_i\, \d C_l
+\int_\mathcal{S} \epsilon_{kjl}\epsilon_{lmn} \sigma_{ij} b_{i,m}\, \d S_n\,.
\end{align}

Since the Burgers vector of a Somigliana dislocation is non-constant,
surface integrals depending on 
the gradient of the Burgers vector appear in Eqs.~(\ref{Mura-S})--(\ref{PK-S})
as characteristic term of a Somigliana dislocation.
If the Burgers vector $\bm b$ is constant, then all the master-equations
for Volterra dislocations follow from Eqs.~(\ref{u-Vol})--(\ref{PK-S}).

\section{Relation to 3D crack modeling based on Somigliana dislocations}

In this section, we give systematically 
the stress fields of 3D cracks based on the stress fields of Somigliana
dislocations. 
Hence, we use the stress fields of Somigliana dislocations
in order to determine the stress field of a 3D crack. 
Such a technique is usually called dislocation based fracture
  mechanics~\citep{W96} or distributed dislocation
  technique~\citep{Hills}. 
Using the stress fields of Somigliana dislocations, 
singular integral equations for 
the stress fields of 3D crack problems can be derived. 
The stress field of a 3D crack can be obtained if one assumes that the crack
surface is filled with a continuous distribution of dislocations.

In this way, the stress field of a Somigliana 
dislocation~(\ref{T-Vol}) gives the stress field of a 3D crack
\begin{align}
\label{T-Vol-Cr}
\sigma_{pq}(\Bx)=\int_\mathcal{S} S_{pqln}(\Bx-\Bx')\, b_l(\Bx')\, \d S'_{n}\,.
\end{align}
In order the crack faces to be stress-free, the stress~(\ref{T-Vol-Cr}) should be
equal and opposite to the traction induced by external loads.
Using the principle of superposition leads to the integral equation
\begin{align}
\label{T-Vol-Cr-B}
\int_\mathcal{S} S_{pqln}(\Bx-\Bx')\, b_l(\Bx')\, \d S'_{n}
=-\sigma^\infty_{pq}(\Bx)\,,
\end{align}
where $\sigma^\infty_{pq}$ denotes an external loading. 
In 3D crack modeling the surface of the Somigliana dislocation loop
$\mathcal{S}$ plays the role of the crack surface, the dislocation line 
of the Somigliana dislocation loop
$\mathcal{S}$ becomes the boundary of the crack surface 
and the ``local'' Burgers vector $\bm b(\Bx')$ 
is the vector of discontinuity in the displacement (or in the opening of the crack)
and plays the role of the dislocation 
distribution function which is to be found from the solution of the 
integral equation~(\ref{T-Vol-Cr-B}). 
If the dislocation distribution function is obtained, then the stress field of
the crack is completely determined.
For anisotropic 3D crack modeling, the anisotropic version of 
the kernel $\bm S$ given in Eq.~(\ref{S-2}) can be used. 
Eq.~(\ref{T-Vol-Cr}) is a hypersingular integral equation possessing $1/R^3$- and
$\delta(\bm R)$-singularities.
An equation like Eq.~(\ref{T-Vol-Cr-B}) was derived by \citet{Kunin} in the 
framework of eigenstrain theory. 
\citet{Hills} proposed an isotropic version of Eq.~(\ref{T-Vol-Cr-B})
in 3D distributed dislocation technique for 3D planar cracks. 
The kernel for planar cracks of arbitrary shape given by \citet{Hills} reads
\begin{align}
\label{K}
K_{pql3}(\Bx-\Bx')=C_{pqim}C_{jkl3} G_{ij,km}(\Bx-\Bx')\,.
\end{align} 
If we compare Eq.~(\ref{K}) with (\ref{S}), we see that in 
Eq.~(\ref{K}) the $\delta(\Bx-\Bx')$-term due to the plastic distortion in the 
additive decomposition~(\ref{B-deco}) is neglected.
Moreover, \citet{Hills} used only the $1/R^3$-part of the kernel~(\ref{K})
for the crack stress field.

Alternatively, the Peach-Koehler stress formula of a Somigliana dislocation
may be used to determine the stress field of a 3D crack
\begin{align}
\label{PKs-S-Cr}
\sigma_{pq}(\Bx)=\oint_\mathcal{C} M_{pqlr}(\Bx-\Bx')\, b_l(\Bx')\, \d C'_{r}
+\int_\mathcal{S} \epsilon_{rst} M_{pqlr}(\Bx-\Bx')\,  b_{l,s'}(\Bx')\, \d S'_{t}
\end{align}
and the following integral equation for the external loading reads
\begin{align}
\label{PKs-S-Cr-B}
\oint_\mathcal{C} M_{pqlr}(\Bx-\Bx')\, b_l(\Bx')\, \d C'_{r}
+\int_\mathcal{S} \epsilon_{rst} M_{pqlr}(\Bx-\Bx')\,  b_{l,s'}(\Bx')\, \d S'_{t}
=-\sigma^\infty_{pq}(\Bx)\,.
\end{align}
It can be seen that Eq.~(\ref{PKs-S-Cr-B}) represents an integral equation 
where also the gradient of the distribution function $b_{l,p'}(\Bx')$ is involved.
For anisotropic 3D crack modeling, the anisotropic version of 
the kernel $\bm M$ is given in Eq.~(\ref{M-2})  
and possessing a $1/R^2$-singularity.

\section{The Eshelby tensor for a spherical inclusion}

In order to give the relation to micromechanics more in detail,
we symmetrize the elastic distortion tensor in Eq.~(\ref{B-1})
to get the elastic strain tensor 
\begin{align}
\label{strain-1}
e_{im}:=\frac{1}{2}\, (\beta_{im}+\beta_{mi})
=-\frac{1}{2}\, C_{jkln} \big(G_{ij,km}+G_{mj,ki}\big)*e^\TP_{ln}
-e_{im}^\TP
\end{align}
with the plastic strain as eigenstrain
$e^\TP_{ln}=1/2(\beta_{ln}^\TP+\beta_{nl}^\TP)$.
For uniform (constant) eigenstrain, Eq.~(\ref{strain-1}) reduces to 
\begin{align}
\label{strain-2}
e_{im}=S^{\text{Esh}}_{imln} e^\TP_{ln}-e_{im}^\TP\,.
\end{align}
The total strain tensor, which is the induced strain of an inclusion
embedded in an infinite medium,
is constant in the inclusion and is related to the eigenstrain by
\begin{align}
e_{im}^\T=S^{\text{Esh}}_{imln} e^\TP_{ln}\,,
\end{align}
where the (interior) Eshelby tensor, which is a constant tensor, is defined by
\begin{align}
\label{Esh}
S^{\text{Esh}}_{imln}
&=-\frac{1}{2}\, 
C_{jkln} \int_V \big(G_{ij,km}(\bm R)+G_{mj,ki}(\bm R)\big)\, \text{d} V'\,.
\end{align}
Thus, the Eshelby tensor connects the total strain with the eigenstrain
and possesses the symmetries
\begin{align}
\label{Esh-symm}
S^{\text{Esh}}_{imln}=S^{\text{Esh}}_{miln}=S^{\text{Esh}}_{imnl}\,.
\end{align}
Using Eq.~(\ref{gradgradG-deco}), it yields \citep{Kroener86}
\begin{align}
E_{ijkm}=-\int_V G_{ij,km}(\bm R)\, \text{d} V'\,
\end{align}
and the Eshelby tensor can be given in terms of the tensor $E_{ijkm}$
\begin{align}
\label{S-E}
S^{\text{Esh}}_{imln}
&=\frac{1}{2}\, 
C_{jkln} \big(E_{ijkm}+E_{mjki}\big)\,.
\end{align}

Using the tensor of elastic constants for an isotropic material 
\begin{align}
\label{C-iso}
C_{jkln}=\mu\bigg(\delta_{jl}\delta_{kn}+\delta_{jn}\delta_{kl}
+\frac{2\nu}{1-2\nu}\, \delta_{jk}\delta_{ln}\bigg)
\end{align}
and
substituting Eq.~(\ref{E-iso}) into Eq.~(\ref{S-E}), 
the interior Eshelby tensor for a spherical inclusion in an isotropic material is obtained as 
(see also~\citet{Mura,Burya,LiS})
\begin{align}
\label{S-iso}
S^{\text{Esh}}_{imln}=
\frac{1}{15(1-\nu)}\, \big[(5\nu-1)\,\delta_{im}\delta_{ln}
+(4-5\nu)(\delta_{il}\delta_{mn}+\delta_{ml}\delta_{in})\big]\,.
\end{align}
On the other hand,
substituting Eq.~(\ref{E-aniso}) into Eq.~(\ref{S-E})
the interior Eshelby tensor for a spherical inclusion in an anisotropic material is obtained as
\begin{align}
\label{S-aniso}
S^{\text{Esh}}_{imln}
=\frac{1}{4\pi}\, C_{jkln}\int_0^{2\pi} 
\big[n_m n_k\,(n C n)_{ij}^{-1}+n_i n_k\,(n C n)_{mj}^{-1}
\big]\,  \text{d} \phi\,.
\end{align}
The anisotropic Eshelby tensor of a sphere~(\ref{S-aniso}) is given as 
line integral around the unit circle
because it is based on the Lifshitz-Rosenzweig-Synge-Barnett representation for the
anisotropic Green tensor~(\ref{G}). Thus, the representation~(\ref{S-aniso}) 
is simpler than the representation 
as integral over the unit sphere as given by~\citet{Kneer} and \citet{Bacon}.
Finally, we conclude that the tensor $E_{ijkm}$ is the tensor solving the 
spherical inclusion problem.

\section{Conclusions}

In this work, we derived the master-equations 
for general dislocation field theory in anisotropic elasticity from the
perspective of micromechanics. 
The general formula, which is the second derivative of the Green tensor, 
is decomposed into a $1/R^3$-term and a Dirac $\delta$-term, 
which is a new and novel formulation. We derived the dislocation
master-equations in general and applied to Somigliana dislocations. 
Moreover, we derived a line integral representation for the interior 
Eshelby tensor (the second order derivative of the Green tensor) 
of a spherical inclusion in anisotropic elastic media, using 
the so-called $\bm F$-tensor. 
The derived dislocation formulation is a contribution 
to dislocation theory in general, which will have potential
impacts to discrete dislocation dynamics, dislocation based fracture mechanics 
and other meso-scale crystal plasticity theories and computations.

\section*{Acknowledgements}
The author gratefully acknowledges a grant from the
Deutsche Forschungsgemeinschaft
(Grant No.~La1974/3-2).
The author thanks Thomas Michelitsch (Paris)
and Helmut Kirchner (Paris) for useful remarks 
during the preparation of this work.

\begin{appendix}
\section{The functions $1/R$, $R$, $R^3$  and their derivatives}
\label{AppendixA}
\setcounter{equation}{0}
\renewcommand{\theequation}{\thesection.\arabic{equation}}

In general, the derivative of order $n$ acting on a Green function 
of a differential operator of order $n$ gives a Dirac $\delta$-term.

We give examples, which are relevant for the present work. \\
{\bf Function $1/R$:}\\
The first derivative reads
\begin{align}
\label{R1}
\Big(\frac{1}{R}\Big){,_i}=-\frac{R_i}{R^3}\, ,
\end{align} 
where $R_i=x_i-x'_i$,
the second derivative is given by 
\begin{align}
\label{R2}
\Big(\frac{1}{R}\Big){,_{ij}}=
-\frac{4\pi}{3}\,\delta(\bm R)\delta_{ij}
-\frac{1}{R^3}\, \delta_{ij}+\frac{3 R_iR_j}{R^5}
\, 
\end{align}
and therefore the trace of Eq.~(\ref{R2}) reads
\begin{align}
\label{R3}
\Big(\frac{1}{R}\Big){,_{ii}}=\Delta\, \frac{1}{R}
=-4\pi\,\delta(\bm R)\,.
\end{align}
Since the function $1/R$ is the Green function of the 
Laplace operator, which is a differential operator of second order,
the derivative of second order acting on $1/R$ generates 
a $\delta(\bm R)$-term.\\
{\bf Function $R$:}\\
The derivatives of $R$ from the first to the fourth order 
are given by the following set of equations
\begin{align}
\label{Ri}
R_{,i}=\frac{R_i}{R}
\end{align} 
\begin{align}
\label{Rij}
R_{,ij}=\frac{\delta_{ij}}{R}-\frac{R_{i}R_j}{R^3}
\end{align} 
\begin{align}
\label{Rii}
R_{,ii}=\Delta R=\frac{2}{R}
\end{align} 
\begin{align}
\label{Rijk}
R_{,ijk}&=-\frac{\delta_{ij}\,R_k+\delta_{ik}\, R_j+\delta_{jk}\, R_i}{R^3}
+\frac{3 R_iR_j R_k}{R^5}
\end{align} 
\begin{align}
\label{Riik}
R_{,iik}&=-\frac{2R_k}{R^3}
\end{align}
\begin{align}
\label{Rijkl}
R_{,ijkl}&=-\frac{8\pi}{15}\,\delta(\bm R)\delta_{ijkl}
-\frac{1}{R^3}\, \delta_{ijkl}
\nonumber\\
&\quad+3\,\frac{\delta_{ij}\,R_kR_l+\delta_{ik}\, R_jR_l+\delta_{il}\, R_jR_k
+\delta_{jk}\,R_iR_l+\delta_{jl}\, R_iR_k+\delta_{kl}\, R_iR_j}{R^5}
\nonumber\\&\quad 
-\frac{15 R_iR_j R_kR_l}{R^7}
\, 
\end{align} 
\begin{align}
\label{Riikl}
R_{,iikl}=-\frac{8\pi}{3}\,\delta(\bm R)\delta_{kl}
-\frac{2}{R^3}\, \delta_{kl}+\frac{6 R_kR_l}{R^5}
\end{align}
\begin{align}
\label{Riikk}
R_{,iikk}=\Delta\Delta R=-8\pi\,\delta(\bm R)
\end{align}
where 
\begin{align}
\label{delta-4-A}
\delta_{ijkm}=\delta_{ij}\delta_{km}+\delta_{ik}\delta_{jm}+\delta_{im}\delta_{jk}\,.
\end{align}
Thus, the function $R$ is the Green function of the 
bi-Laplace operator, which is a differential operator of fourth order,
and every derivative of fourth order acting on $R$ generates 
a $\delta(\bm R)$-term.\\
{\bf Function $R^3$:}\\
The derivatives of $R^3$ from the first to the fourth order 
are given by the following set of equations
\begin{align}
\label{R3i}
R^3_{,i}=3R_i R 
\end{align} 
\begin{align}
\label{R3ij}
R^3_{,ij}=3 \delta_{ij}R+\frac{3R_{i}R_j}{R}
\end{align} 
\begin{align}
\label{R3ii}
R^3_{,ii}=\Delta R^3=12 R
\end{align} 
\begin{align}
\label{R3ijk}
R^3_{,ijk}=3\,\frac{\delta_{ij}\,R_k+\delta_{ik}\, R_j+\delta_{jk}\, R_i}{R}
-\frac{3 R_iR_j R_k}{R^3}
\end{align} 
\begin{align}
\label{R3iik}
R^3_{,iik}&=\frac{12R_k}{R}
\end{align}
\begin{align}
\label{R3ijkl}
R^3_{,ijkl}&=
\frac{3}{R}\, \delta_{ijkl}
-3\,\frac{\delta_{ij}\,R_kR_l+\delta_{ik}\, R_jR_l+\delta_{il}\, R_jR_k
+\delta_{jk}\,R_iR_l+\delta_{jl}\, R_iR_k+\delta_{kl}\, R_iR_j}{R^3}
\nonumber\\&\quad 
+\frac{9 R_iR_j R_kR_l}{R^5}
\, 
\end{align} 
\begin{align}
\label{R3iikl}
R^3_{,iikl}=\frac{12}{R}\, \delta_{kl}-\frac{12 R_kR_l}{R^3}
\end{align}

\begin{align}
\label{R3iikk}
R^3_{,iikk}=\Delta\Delta R^3=\frac{24}{R}\,.
\end{align}
Combining Eq.~(\ref{R3iikk}) with Eq.~(\ref{R3}), it can be seen that
$R^3$ is the Green function of the tri-Laplace operator, which is a
differential operator of sixth order 
\begin{align}
\label{R3iikkmm}
R^3_{,iikkmm}=\Delta\Delta\Delta R^3= -24\cdot 4\pi\,\delta(\bm R)\,.
\end{align}
Thus, the derivative of sixth order acting on $R^3$ generates 
a $\delta(\bm R)$-term.

\section{Useful relations}
\label{AppendixB}
\setcounter{equation}{0}
\renewcommand{\theequation}{\thesection.\arabic{equation}}

First, we prove the relation:
\begin{align}
\label{relB-1}
\tau_l \pd_l \big(n_m n_k\,(n C n)_{ij}^{-1}\big)
=\tau_l \big(n_k n_{m,l}+n_m n_{k,l} \big)(n C n)_{ij}^{-1}
+n_kn_m \tau_l (n C n)_{ij,l}^{-1}=0\,.
\end{align}
Using 
\begin{align}
\label{relB-1-1}
\tau_m n_m=0
\end{align}
and 
\begin{align}
\label{relB-1-2}
\tau_{m,l}=\frac{1}{R}\,\big(\delta_{lm}-\tau_l\tau_m\big)\,,
\end{align}
we obtain
\begin{align}
\label{relB-1-3}
\pd_l (\tau_m n_m)
&=\tau_{m,l} n_m+\tau_m n_{m,l}
=\frac{n_l}{R} +\tau_m n_{m,l}
=\Big(\frac{1}{R}\,n_l\tau_m +n_{m,l}\Big)\tau_m
=0\,,
\end{align}
since $\tau^2=1$. 
Thus, if $\tau_m\neq 0$, we have
\begin{align}
\label{relB-1-4}
n_{m,l}=-\frac{1}{R}\, n_l \tau_m\,.
\end{align}
Moreover, it yields
\begin{align}
\label{relB-1-5}
(n C n)_{ij,l}^{-1}=\frac{n_l}{R}\, F_{ij}
\end{align}
with 
\begin{align}
\label{F-ij-B}
F_{ij}=(n C n)_{ip}^{-1}\big[(n C \tau)_{pq}+(\tau C n)_{pq}\big](n C
n)_{qj}^{-1}\,,
\end{align}
then Eq.~(\ref{relB-1}) is fulfilled.

Second, we prove the relation:
\begin{align}
\label{relB-2-B} 
\Delta \big(n_m n_k\,(n C n)_{ij}^{-1}\big)
=\frac{1}{R^2} \Big(
2\tau_m \tau_k\,(n C n)_{ij}^{-1}
-2 (n_m \tau_k+n_k\tau_m) F_{ij}
+n_m n_k A_{ij}\Big)\,,
\end{align}
where
\begin{align}
\label{A-ij-B}
A_{ij}&=F_{ip}\big[(n C \tau)_{pq}+(\tau C n)_{pq}\big](n C n)_{qj}^{-1}
+(n C n)_{ip}^{-1}\big[(n C \tau)_{pq}+(\tau C n)_{pq}\big]F_{qj}
\nonumber\\
&\quad 
-2 (n C n)_{ip}^{-1}(\tau C \tau)_{pq}(n Cn)_{qj}^{-1}\,.
\end{align}
We calculate
\begin{align}
\label{relB-3-B}
\pd_l \pd_l \big(n_m n_k\,(n C n)_{ij}^{-1}\big)
&=\big(n_k n_{m,ll}+n_m n_{k,ll}+2n_{k,l}n_{m,l} \big)(n C n)_{ij}^{-1}\nonumber\\
&\quad
+2\big(n_k n_{m,l}+n_m n_{k,l} \big)(n C n)_{ij,l}^{-1}
+n_kn_m  (n C n)_{ij,ll}^{-1}\,.
\end{align}
Using Eqs.~(\ref{relB-1-1}), (\ref{relB-1-2}), (\ref{relB-1-4})--(\ref{F-ij-B}), 
the relations
\begin{align}
n_{m,ll}=-\frac{n_m}{R^2}
\end{align}
and 
\begin{align}
(n C n)_{ij,ll}^{-1}=\frac{n_l}{R} F_{ij,l}=\frac{1}{R^2}\, S_{ij}
\end{align}
where
\begin{align}
\label{S-ij}
S_{ij}&=F_{ip}\big[(n C \tau)_{pq}+(\tau C n)_{pq}\big](n C n)_{qj}^{-1}
+(n C n)_{ip}^{-1}\big[(n C \tau)_{pq}+(\tau C n)_{pq}\big]F_{qj}
\nonumber\\
&\quad 
+2 (n C n)_{ip}^{-1}
\big[(n C n)_{pq}-(\tau C \tau)_{pq}\big](n Cn)_{qj}^{-1}\,.
\end{align}
with  the relation
\begin{align}
(n C n)_{ip}^{-1} (n C n)_{pq} (nCn)_{qj}^{-1}
=(n C n)_{ij}^{-1}
\end{align}
we have
\begin{align}
\label{A-ij-B2}
A_{ij}&=S_{ij}-2 (nCn)_{ij}^{-1}\,. 
\end{align}
Finally, Eq.~(\ref{relB-2-B}) is obtained from Eq.~(\ref{relB-3-B}). 

\end{appendix}

\end{document}